# Photo-Activated Colloidal Dockers for Cargo Transportation

Jérémie Palacci,* Stefano Sacanna, Adrian Vatchinsky, Paul M. Chaikin, and David J. Pine

Center for Soft Matter Research, Department of Physics, New York University, USA

*Supporting Information Placeholder*

**ABSTRACT:** We introduce a self-propelled colloidal hematite docker that can be steered to a small particle cargo many times its size, dock, transport the cargo to a remote location, and then release it. The self-propulsion and docking are reversible and activated by visible light. The docker can be steered either by a weak uniform magnetic field or by nanoscale tracks in a textured substrate. The light-activated motion and docking originate from osmotic/phoretic particle transport in a concentration gradient of fuel, hydrogen peroxide, induced by the photo-catalytic activity of the hematite. The docking mechanism is versatile and can be applied to various materials and shapes. The hematite dockers are simple single-component particles and are synthesized in bulk quantities. This system opens up new possibilities for designing complex micron-size factories as well as new biomimetic systems.

Controlled motion and transport of objects are basic functions that are simple to perform at the macro-scale and indispensable for manufacturing and robotics. At the micro-scale, synthetic agents performing these tasks would be very useful for biomedical applications such as drug-delivery, in situ assembly, delivery of microscopic devices, and for microfluidics. However, these tasks are difficult at small length-scales where reversible and wireless actuation remain a significant challenge. This has fueled a significant effort to design populations of artificial micro-agents capable of moving autonomously in a controlled fashion while performing complex tasks[1-3].

One of the key requirements for building a micro-robot is the ability of the system to harvest the free energy from its environment and convert it into mechanical work. The energy source can be provided by an electromagnetic field[4,5] or by chemical fuels. Different routes and mechanisms have been explored for the latter: (i) jet propulsion of microtubular engines or (ii) self-phoretic particles. In the first case, the fuel is catalytically transformed into gas bubbles and expelled, propelling particles to ultrafast speeds, e.g. 350 body lengths/s[6-8]. Self-phoretic propulsion relies on an interfacial phenomenon, phoresis, which leads to migration of a colloid in some kind of gradient[9]. Self-electrophoresis was first used to propel bimetallic nanorods[10] in hydrogen peroxide in work by Paxton and Sen[11] and has been broadly studied since[1,2]. Alternatively, many realizations of microrobots are based on self-diffusiophoresis, the autonomous motion of a colloid in a chemical gradient produced by the anisotropic chemical activity of the particle[12]. This route has been used to produce self-propelled Janus colloids in hydrogen peroxide[13] and micromotors in diverse media and chemicals[14-17]. The development of the micromotors has been carried out simultaneously with efforts to enrich their range of function, for example for steering[18], fuel free locomotion[19], light activation[20], cargo[21,22] or for transport of cells[23], emulsion droplets[24], or colloids in a microfludic environment,[18,22,25-30] and for self-assembly and other collective effects[31,32]. Other functionalities were recently realized with the development of self-propelled nanotools[6], which are useful for biomedical applications. More information about the mechanisms and experimental realizations can be found in recent reviews[1-3].

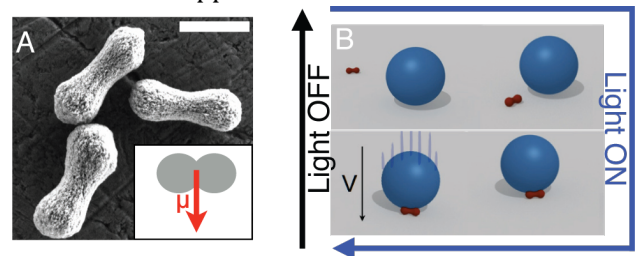

**Figure 1.** (A) SEM picture of the hematite peanut particles. Scale bar is 1 μm. Inset shows the direction of the permanent magnetic moment μ of the particle, perpendicular to the long axis. (B) Schematic of docking. Under activation by an external blue light and steering by a magnetic field, a hematite particle docks with a passive sphere and transports it. This scenario is reversible: when the light is turned off, the hematite particles release the cargo and diffuse away.

These self-propelled particles are obtained with various techniques including "rolled-up" technologies to obtain microtubes[33,34] or vapor-deposition on colloids, limited by the two-dimension nature of the process. Moreover, their synthesis is generally complex, being composed of a number of iterative steps, at least one for each additional functionality. For example, one layer of magnetic material is required for direction control, another layer for docking, and a final layer of a chemically active material to provide propulsion. Here we present a novel type of particle made from hematite, a photo-catalytic iron oxide. These particles are synthesized in very large quantities, and provide all the desired functionalities: they self-propel, dock and release particles with light actuation, and can be externally steered by a weak magnetic field.

Our particles are made from hematite, a canted antiferromagnetic material[35] with a permanent magnetic moment μ. The particles are synthesized in bulk, and can routinely be synthesized in 10 ml suspensions containing 20%v/v. Various shapes can be obtained: cubes, ellipses or peanuts, in sizes ranging from tens of nanometers to a few microns[36]. In this paper, we focus on the case of "peanut-shaped" hematite colloids, typically ~1.5μm long and ~0.6μm wide, [Figure 1A]. The permanent magnetic moment μ is directed perpendicular to the long axis, and can be deduced by direct optical observation [Figure 1A.-inset]. To improve the performance of the peanut particles as dockers, the hematite surface is partially etched using hydrochloric acid (HCl) solutions (usually 5 M). This creates particles with a rough surface that responds better to light than unetched particles. This effect is discussed later along with the propulsion mechanism. The particles can be observed with an optical microscope and are immersed in a basic solution (pH~8.5) containing hydrogen peroxide (1% w/w), 5 mM tetramethylammonium hydroxide (TMAH) in deionized water.

The colloids sediment under gravity and reside near the surface of a glass capillary. They are at equilibrium with the solvent and exhibit thermal Brownian motion. When illuminated through a microscope objective (100×, N.A.=1.4) with blue light (Nikon Intensilight, filtered within λ~430-490nm), the particles are attracted to the surface and start propelling along the surface. A weak uniform magnetic field $B$~1 mT can be externally applied to the sample using a Helmoltz coil. It fixes the orientation of the magnetic moment μ, thus freezing the rotational diffusion of the peanut. Under light-activation, the particle self-propels along the direction of $B$. Rotating the direction of the field induces a magnetic torque and a rapid change of direction in the self-propulsion, allowing

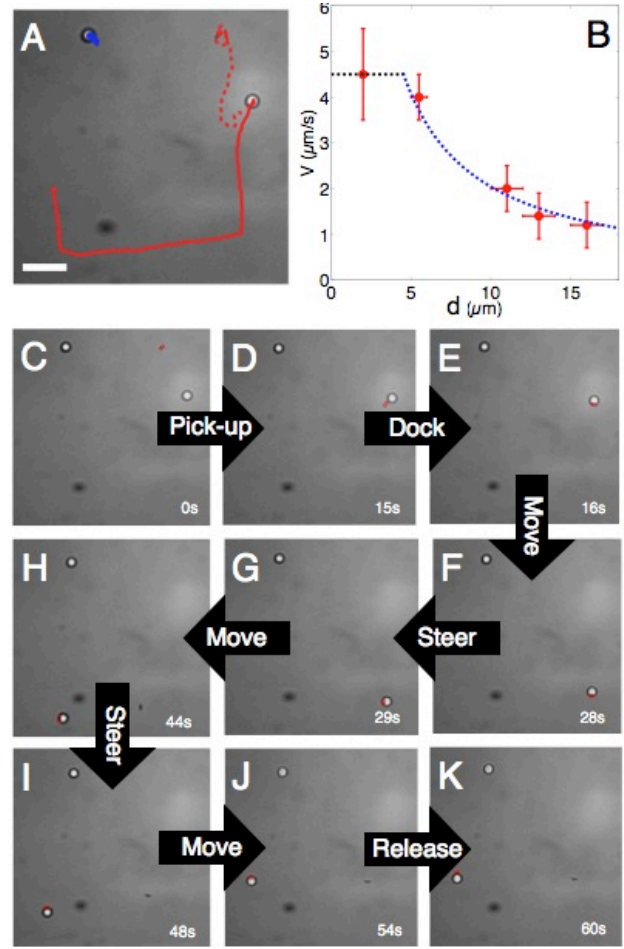

**Figure 2.** Photo activated colloidal dockers (A) Trajectory of a hematite particle steered (dashed red line) using an external magnetic field to pick up a 5 μm colloidal cargo together with its trajectory after docking (solid red line). The blue trace shows the Brownian diffusion of a control colloid. (B) Dependence of the velocity $V$ on the diameter d of the cargo. (C-K) Time lapse of the colloidal cargo experiment. The hematite particle position is indicated by red boxes. (C-D) At $t$=0s, the hematite particle is activated by light. At $t$=6 s, an external magnetic field is applied to steer the hematite particle towards a 5 μm colloid, the cargo. The hematite particle moves along the direction of the magnetic field. (E-F and H-I) The hematite particle docks on the colloid and tows it, with the hematite peanut leading. The direction is fixed externally by the magnetic field. (F-G and I-J) Changing the direction of the magnetic field steers the hematite peanut, quickly redirecting the cargo. (J-K). The light is switched off, the phoretic attraction ceases, and the colloidal cargo is released. The hematite peanut and cargo resume Brownian diffusion.

one to steer the particles, as shown in [Figure 2C-D].

The mechanism of self-propulsion has been discussed in a previous publication[37]. Briefly, exposing the hematite peanuts to blue light induces the photocatalytic decomposition of the hydrogen peroxide in solution, $2H_2O_2 \rightarrow O_2 + 2H_2O$. This establishes chemical gradients in the vicinity of the peanut particles thus depleting $H_2O_2$ and creating an excess of $O_2$. In a chemical gradient, there is an unbalanced osmotic pressure in the interfacial layer near any nearby surface. The unbalanced osmotic pressure induces an interfacial diffusio-osmotic flow along the substrate, which initially propels the particle towards the substrate, in this case the capillary cell wall. Once against the wall, the gradient is in principle symmetric along the wall and the particle should remain still. This is what we observe for most unetched peanut particles. After etching, however, about 80% of the peanuts start to self-propel along the substrate. We attribute this effect to the enhanced chemical anisotropy of the peanut surfaces after roughening. The peanut particles preferentially propel in a direction perpendicular to their long axis.

The illuminated hematite harvests free energy from the hydrogen peroxide fuel in solution generating an osmotic flow along the substrate. A consequence of this peculiar self-propulsion mechanism is a sensitivity of the active particles to the physical properties of the substrate. This sensitivity can be harnessed to direct the particles along the nanometer-size tracks in a textured substrate. We defer discussion of this phenomenon to the final section.

The chemical gradients surrounding the activated hematite induce diffusiophoresis of surrounding colloids in the solution. For solutions containing TMAH (pH~ 8.5), we observe negative phoresis (attraction of a sphere towards the hematite particle) for all materials tested: silica, polystyrene and 3-methacryloxypropyl trimethoxysilane (TPM). Alternatively, lowering the pH to 6.5, suppressing the TMAH from the solution, we observe positive diffusiophoresis: spheres are repelled from the hematite particle. In the following, we exploit only the ability to attract colloidal spheres to the hematite, and hence consider only basic solutions (pH~8.5) containing TMAH. We use this property to dock the hematite particles to larger colloids and to carry them as cargo, as sketched in [Figure 1B] and documented in Figure 2A [also see Movie 1].

First we activate the particles with light. Using a weak uniform external magnetic field $B \sim 1$ mT, we then direct the peanut particle to the vicinity of a large colloid [Figure 2C-D]. The hematite particle phoretically attracts the large colloid and docks on its surface [Figure 2E]. The composite hematite-peanut/colloid system forms an asymmetric particle with a localized chemically active site, the hematite peanut, and a passive part, the sphere. The system propels as a whole, with the active hematite peanut leading [Figure 2E-F]. The direction of the cargo is fixed with the external magnetic field using the hematite particle to steer [Figure 2F-G]. Turning the light off, the chemical activity of the hematite ceases and the chemical concentration gradients vanish by diffusion in a few tens of milliseconds. In the absence of gradients, the osmotic motion and phoretic attraction cease as the system returns to equilibrium. As the peanut particles stops, the colloidal cargo is released and diffuses away [Figure 2J-K].

The docking mechanism is reversible: attraction and propulsion immediately restart once the light is turned on. Consequently, a hematite particle can carry, dock and release many cargos and, for example, play the role of a microscopic shepherd gathering colloidal particles [movie 2]. The effect is versatile; we can load any particle exhibiting negative diffusiophoresis, in our case silica, polystyrene, and TPM for diameters ranging from 1 to 20 μm.

While the velocity of isolated hematite particles is widely distributed, the velocities of composite hematite/cargo pairs are all the same for a given cargo size. In this case, the chemical gradient propelling a composite hematite/cargo pair is determined by the geometrical anisotropy of the pair, not by the chemical anisotropy of the etched lone carrier. Indeed, once docked, even previously immobile unetched hematite particles start propelling.

We measure the velocity of the composite hematite/colloid system while varying the diameter $d$ of the cargo [Figure 2B]. For cargos significantly larger than the hematite carrier ($d > 5$ μm), the transport velocity $V$ scales inversely with the cargo diameter, $V \propto 1/d$, consistent with a constant pulling force exerted by the hematite particle and a Stokes drag proportional to $d$. For cargos comparable in size to the hematite carrier, the velocity saturates.

The cargo scenario is readily adapted to other hematite particle shapes. We can dock colloidal particles using hematite cubes, with the minor disadvantage that the direction of the magnetic moment, which is along the body diagonal of the cube, is not apparent by optical observation.

The interfacial origin of the propulsion mechanism makes the particles sensitive to chemical or physical alterations of the substrate. As a demonstration, we prepared a textured substrate of parallel stripes ob-tained by making a polydimethylsiloxane (PDMS) replica of a compact disk (CD)[38]. The pattern consists of stripes that are 0.6 μm wide, 110 nm deep, and separated by 1.2 μm grooves [Figure 3A]. The PDMS replica exhibits visible channels [Figure 3B and Figure 3C-F]. Before turning on the light, the hematite and a 5 μm TPM sphere diffuse near the substrate surface, unaffected by the shallow pattern [Figure 3C]. After turning on the light, the hematite aligns along a channel and phoretically attracts the TPM sphere [Figure 3D]. The loading of the colloid breaks the symmetry and propels the cargo along the lines of the pattern [Figure 3E,F and movie 3] showing that the hematite particles can be driven along a predetermined pathway using physical alteration of the landscape. The same effect is observed on a cleaved mica surface where the hematite particles self-propel along a nano-crack in the material. In both cases, the cargo is released by turning off the light.

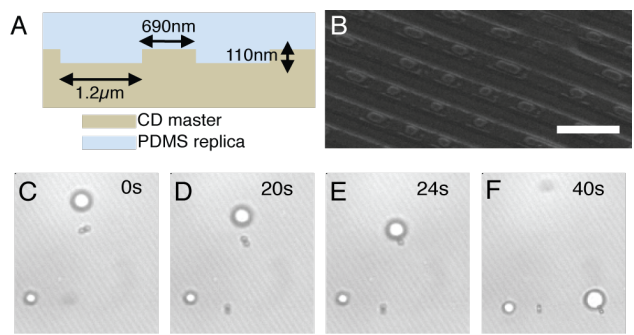

**Figure 3.** (A) A textured PDMS substrate of parallel lines is made using a CD master. (B) SEM micrograph of the PDMS replica. The scale bar is 2 μm. (C) The hematite and large-cargo colloids sediment near the substrate pattern and undergo Brownian diffusion. (D) When illuminated, the hematite is attracted to the substrate and aligns with the texture. (E-F) The colloid cargo is attracted to the hematite and docks. The composite hematite and colloid cargo then start moving, with the hematite in front and following the tracks imprinted in the substrate. No external field is applied to steer the particles.

In this paper, we have introduced a scheme for making micro-robots, synthesized in bulk, possessing the ability to be activated by light, and steered by a magnetic field. They can load, transport, and unload colloids made from many different materials, with sizes up to 20 μm, many times the size of the micro-robots. These capabilities open up new opportunities for engineering at the microscale and for micromanufacturing. We demonstrate their ability to act as a colloidal shepherd to pick up and gather spheres. This is a step forward in making a large-scale microscopic factory. Moreover, the ability to autonomously carry the cargo along a pre-determined pathway drawn on a textured substrate is unique and a direct translation of ground rail transportation to the microscale. In this sense, the system acts similarly to molecular motors like myosin, which walk along one dimensional actin filaments[39]. The interaction between myosin and actin is at the core of transport in cells, cell division, or muscle contraction. The ability to produce synthetic systems inspired by motor proteins such as myosin opens up new possibilities for complex biomimetic systems such as microscopic artificial muscles.

## ASSOCIATED CONTENT

**Supporting Information.** Materials and Methods, Supplementary movies. This material is available free of charge via the Internet at http://pubs.acs.org.

## AUTHOR INFORMATION
**Corresponding Author**
jp153@nyu.edu

**Notes**

The authors declare no competing financial interest.

## ACKNOWLEDGMENT

This work was supported by the MRSEC Program of the Na-tional Science Foundation under Award Number DMR-0820341 and by the U. S. Army Research Office under Grant Award No. W911NF-10-1-0518.

Table of Contents artwork

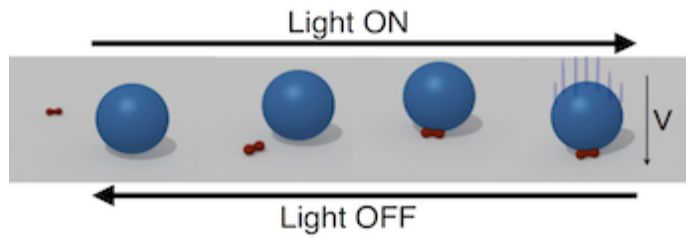